\documentclass[pra,twocolumn,groupedaddress,showpacs,floatfix]{revtex4}

\usepackage{graphics}
\usepackage{graphicx}
\usepackage{bm}
\usepackage{amsmath}
\usepackage{amsfonts}
\usepackage{amssymb}
\usepackage{latexsym}

\begin{document}

\title{Multiphoton entanglement through a Bell multiport beam splitter} 
\author{Yuan Liang Lim$^1$\footnote{Email: yuan.lim@imperial.ac.uk} and Almut Beige$^{2,1}$\footnote{Email: a.beige@imperial.ac.uk}}
\affiliation{$^1$Blackett Laboratory, Imperial College London, Prince Consort Road, London SW7 2BZ, United Kingdom\\
$^2$Department of Applied Mathematics and Theoretical Physics, University of Cambridge, \\ Wilberforce Road, Cambridge CB3 0WA, United Kingdom}

\date{\today}

\begin{abstract}
Multiphoton entanglement is an important resource for linear optics quantum computing. Here we show that a wide range of highly entangled multiphoton states, including {\em W}-states, can be prepared by interfering  {\em single} photons inside a Bell multiport beam splitter and using postselection. A successful state preparation is indicated by the collection of one photon per output port. An advantage of the Bell multiport beam splitter is that it redirects the photons without changing their inner degrees of freedom. The described setup can therefore be used to generate polarisation, time-bin and frequency multiphoton entanglement, even when using only a single photon source.
\end{abstract}

\maketitle

\section{Introduction} \label{Introduction}

Entanglement spurs a great deal of interest in quantum information processing \cite{Gottesman,Knill}, quantum cryptography \cite{Bennett,Ekert} and for fundamental tests of quantum mechanics \cite{Bell,Aspect}.  For many practical purposes, photons provide the most favoured qubits as they possess very long lifetimes and an ease in distribution. However, it is not possible to create a direct interaction between photons and hence they are difficult to entangle. One way to overcome this problem is to create polarisation or time-bin entanglement via photon pair creation within the same source as in atomic cascade and parametric down-conversion experiments. This has already been demonstrated experimentally by many groups \cite{Aspect,downconversion,Gisin}. Other, still theoretical proposals employ certain features of the combined level structure of atom-cavity systems \cite{Gheri,Lange,schoen}, photon emission from atoms in free space \cite{jpa} or accordingly initialised distant single photon sources \cite{orlando,moonlight}.

Alternatively, highly entangled multiphoton states can be prepared using independently generated single photons with no entanglement in the initial state, linear optics and postselection. In general, the photons should enter the linear optics network such that the information about the origin of each photon is erased. Afterwards postselective measurements are performed in the output ports of the network \cite{Sipe}. Using this approach, Shih and Alley verified the generation of maximally entangled photon pairs in 1988 by passing two photons simultaneously through a 50:50 beam splitter and detecting them in different output ports of the setup \cite{Shih}. For a recent experiment based on this idea using quantum dot technology, see Ref.~\cite{recent}.

Currently, many groups experimenting with single photons favour parametric down conversion because of the quality of the produced output states. However, these experiments cannot be scaled up easily, since they do not provide efficient control over the arrival times of the emitted photons. It is therefore experimentally challenging to interfere more than two photons successfully. Interesting experiments involving four photons have nevertheless been performed \cite{Weinfurter,pan} but going to higher photon numbers might require different technologies. To find alternatives to parametric down conversion, a lot of effort has been made over the last years to develop sources for the generation of single photons on demand \cite{Law,Kuhn1,Duan,Kwiat}. Following these proposals, a variety of experiments has already been performed, demonstrating the feasibility and characterising the quality of these sources based on atom-cavity systems \cite{Kuhn2,Mackeever,Lange04}, quantum dots \cite{Yamamoto} and NV color centres \cite{Weinfurter00,Grangier}. 

Motivated by these recent developments, several authors studied the creation of multiphoton entanglement by passing photons generated by a single photon source through a linear optics network \cite{Zukowski,Kok,Fiurasek,Xubo,Wang,Sagi,Pryde}. A variety of setups has been considered. Zukowski {\em et al.} showed that the $N \times N$ Bell multiport beam splitter can be used to produce higher dimensional EPR correlations \cite{Zukowski}. Special attention has been paid to the optimisation of schemes for the generation of the so-called NOON state with special applications in lithography \cite{Kok,Fiurasek,Xubo,Pryde}. Wang studied the event-ready generation of maximally entangled photon pairs without number resolving detectors \cite{Wang} and Sagi proposed a scheme for the generation of $N$-photon polarisation entangled GHZ states \cite{Sagi}. It is possible to prepare arbitrary multiphoton states \cite{Cerf} using for example probabilistic but universal linear optics quantum gates, like the one described in Ref.~\cite{Franson}, but this approach is not always the most favourable. 

\begin{figure}
\begin{minipage}{\columnwidth}
\begin{center}
\resizebox{\columnwidth}{!}{\rotatebox{0}{\includegraphics{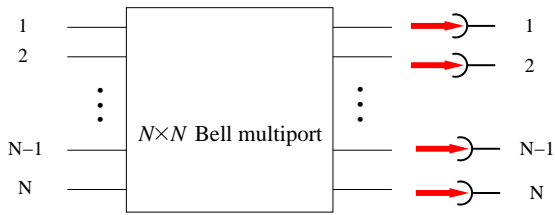}}}  
\end{center}
\caption{Experimental setup for the generation of multiphoton entanglement by passing $N$ single photons through an $N \times N$ Bell multiport beam splitter. The state preparation is considered successful under the condition of the collection of one photon per output.} \label{scheme}
\end{minipage}
\end{figure}

Here we are interested in the generation of highly entangled qubit states of $N$ photons using only a single photon source and a symmetric $N \times N$ Bell multiport beam splitter, which can be realised by combining single beam splitters into a symmetric linear optics network with $N$ input and $N$ output ports \cite{Zukowski,Paivi1}. In the two-photon case, the described scheme simplifies to the experiment by Shih and Alley \cite{Shih}. To entangle $N$ photons, every input port $i$ of the Bell multiport should be entered by a single photon prepared in a state $|\lambda_i \rangle$. The photons then interfere with each other before leaving the setup (see Fig.~\ref{scheme}). We consider the state preparation as successful under the condition of the collection of one photon per output port, which can be relatively easily distinguished from cases with at least one empty output port. 

One advantage of using a Bell multiport beam splitter for the generation of multiphoton entanglement is that it redirects the photons without changing their inner degrees of freedom, like polarisation, arrival time and frequency. The described setup can therefore be used to generate polarisation, time-bin and frequency entanglement. Especially, time-bin entanglement can be very robust against decoherence and has, for example, applications in long-distance fibre communication \cite{tb}. Moreover, the preparation of the input product state does not require control over the relative phases of the incoming photons, since the phase factor of each photon contributes at most to a global phase of the combined state with no physical consequences. 

This paper is organised as follows. In Section \ref{scatter} we introduce the notation for the description of photon scattering through a linear optics setup. Section \ref{fourphoton} shows that a wide range of highly entangled photon states can be obtained for $N=4$, including the {\em W}-state, the GHZ-state and a double singlet state. Afterwards we discuss the generation of $W$-states for arbitrary photon numbers $N$ and calculate the corresponding probabilities for a successful state preparation. Finally we conclude our results in Section \ref{conclusions}.

\section{Photon scattering through a linear optics setup} \label{scatter}

Let us first introduce the notation for the description of the transition of the photons through the $N \times N$ Bell multiport beam splitter. In the following, $|+ \rangle$ and $|- \rangle$ are the state of a photon with polarisation ``$+$'' or ``$-$'' respectively.  Alternatively, $|+ \rangle$ could describe a single photon with an earlier arrival time or a higher frequency than a photon prepared in $|- \rangle$. As long as the states $|\pm \rangle$ are orthogonal and the incoming photons are in the same state with respect to all other degrees of freedom, the calculations presented in this paper apply throughout. Moreover, we assume that each input port $i$ is entered by one independently generated photon prepared in $|\lambda_i \rangle = \alpha_{i+} |+\rangle_i + \alpha_{i-} |-\rangle_i$, where $\alpha_{i \pm}$ are complex coefficients with $|\alpha_{i+} |^2 + |\alpha_{i-}|^2 = 1$. If $a_{i \mu}^\dagger$ denotes the creation operator for one photon with  parameter $\mu$ in input port $i$, the $N$-photon input state can be written as 
\begin{eqnarray} \label{in}
|\phi_{\rm in} \rangle &=& \prod_{i=1}^N \Big( \sum_{\mu=+,-} 
\alpha_{i \mu} \, a_{i \mu}^\dagger \Big) \, |0 \rangle 
\end{eqnarray}
with $|0 \rangle$ being the vacuum state with no photons in the setup. 

Let us now introduce the unitary $N \times N$-multiport transformation operator, namely the scattering matrix $S$, that relates the input state of the system to the corresponding output state
\begin{eqnarray} \label{fin}
|\phi_{\rm out} \rangle &=& S \, |\phi_{\rm in} \rangle \, .
\end{eqnarray}
Using Eq.~(\ref{in}) and the relation $S^\dagger S = I \!\! I$ yields
\begin{eqnarray} \label{fin2}
|\phi_{\rm out} \rangle &=& S \, \Big( \sum_{\mu=+,-} \alpha_{1 \mu} \, 
a_{1 \mu}^\dagger \Big) 
\, S^\dagger S \, \Big( \sum_{\mu=+,-} \alpha_{2 \mu} \, a_{2 
\mu}^\dagger \Big) \nonumber \\
&& \cdot \, . \, . \, . \, \cdot S^\dagger S \, \Big( \sum_{\mu=+,-} 
\alpha_{N \mu} \, 
a_{N \mu}^\dagger \Big) \, S^\dagger S  \, |0 \rangle \nonumber \\
&=& \prod_{i=1}^N \, \Big( \, \sum_{\mu=+,-} \alpha_{i \mu} \, S \, a_{i 
\mu}^\dagger \, S^\dagger \, \Big) \, |0 \rangle \, .
\end{eqnarray}
In the following, the matrix elements $U_{ji}$ of the unitary transformation matrix $U$ denote the amplitudes for the  redirection of a photon in input $i$ to output $j$. Since the multiport beam splitter does not contain any elements that change the inner degrees of freedom of the incoming photons, the transition matrix $U$ does not depend on $\mu$. Denoting the creation operator for a single photon with parameter $\mu$ in output port $j$ by $b_{j \mu}^\dagger$ therefore yields 
\begin{eqnarray} \label{tran}
S \, a_{i \mu}^\dagger \, S^\dagger = \sum_j U_{ji} \, b_{j \mu}^\dagger \, .
\end{eqnarray}
Inserting this into Eq.~(\ref{fin}) we can now calculate the output state of the $N \times N$ multiport given the initial state (\ref{in}) and obtain
\begin{eqnarray} \label{output1}
|\phi_{\rm out} \rangle &=& \prod_{i=1}^N \, \Bigg[ \, \sum_{j=1}^N \, U_{ji} \, 
\Big( \sum_{\mu=+,-} \alpha_{i \mu} \, b_{j \mu}^\dagger \Big) \, 
\Bigg] \, |0 \rangle \, . ~
\end{eqnarray}
This equation describes the independent redirection of all photons to their potential output ports. Conservation of the norm of the state vector is guaranteed by the unitarity of the transition matrix $U$.

The state preparation is considered successful under the condition of the collection of one photon per output port. To calculate the final state, we apply the corresponding projector to the output state (\ref{output1}) and find that the thereby postselected $N$-photon state equals, up to normalisation, 
\begin{equation} \label{output2}
|\phi_{\rm pro} \rangle=\sum_{\sigma} \Bigg[ \prod_{i=1}^N U_{\sigma(i) i} \Big( 
\sum_{\mu=+,-} \alpha_{i \mu} b_{\sigma (i) \mu}^{\dagger} \Big) 
\Bigg]  \, |0 \rangle \, .~
\end{equation}
Here $\sigma$ are the $N!$ possible permutations of the $N$ items $\{1,\, 2, \, ..., \, N\}$. Moreover, the norm of the state (\ref{output2}) squared, namely
\begin{equation} \label{suc}
P_{\rm suc} = \| \, |\phi_{\rm pro} \rangle \, \|^2 \, ,
\end{equation}
is the success rate of the scheme and probability for the collection of one photon in each output $j$.

\subsection{The Bell multiport beam splitter}

Motivated by a great variety of applications, we are particularly interested in the generation of highly entangled photon states of a high symmetry, an example being {\em W}-states. This suggests to consider symmetric multiports, which redirect each incoming photon with equal probability to all potential output ports. A special example for such an $N \times N$ multiport is the Bell multiport beam splitter. Its transformation matrix
\begin{eqnarray}\label{fourier}
U_{ji} &=& {\textstyle{1 \over \sqrt{N}} \, \omega_N^{(j-1)(i-1)}}
\end{eqnarray} 
is also known as a discrete fourier transform matrix and has been widely considered in the literature \cite{Zukowski,Paivi1,Paivi2}. Here $\omega_N$ denotes the $N$-th root of unity,
\begin{equation} \label{root}
\omega_N \equiv \exp \left( 2{\rm i} \pi /N \right) \, .
\end{equation} 
Proceeding as in Section II.D of Ref.~\cite{Zukowski}, it can easily be verified that $U$ is unitary as well as symmetric. Especially for $N=2$, the transition matrix (\ref{fourier}) describes a single 50:50 beam splitter.  

\section{The generation of 4-photon states} \label{fourphoton}

Before we discuss the $N$-photon case, we investigate the possibility to prepare highly entangled 4-photon states using specially prepared photons and a $4 \times 4$ Bell multiport beam splitter. For $N=4$, the transition matrix (\ref{fourier}) becomes 
\begin{eqnarray} \label{matrix4}
U &=& {\textstyle {1 \over 2}} \left( \begin{array}{rrrr} 
1 & 1 & 1 & 1\\
1 & \omega_4 & \omega_4^2 & \omega_4^3\\
1 & \omega_4^2 & \omega_4^4 & \omega_4^6\\
1 & \omega_4^3 & \omega_4^6 & \omega_4^9 \end{array} \right)  
= {\textstyle {1 \over 2}} \left( \begin{array}{rrrr}
1 & 1 & 1 & 1\\
1 & {\rm i} & -1 & -{\rm i}\\
1 & -1 & 1 & -1\\
1 & -{\rm i} & -1 & {\rm i} \end{array} \right) \, .~  \nonumber \\ &&
\end{eqnarray}
The following analysis illustrates the richness of the problem as well as it motivates possible generalisations for the case of arbitrary photon numbers. 

\subsection{Impossible output states} \label{pose}

Let us first look at the seemingly trivial situation, where every input port of the multiport beamspliter is entered by one photon in the same state, let us say in $|+ \rangle$, so that
\begin{equation} \label{inni}
|\phi_{\rm in} \rangle = a_{1 +}^{\dagger}a_{2 +}^{\dagger}a_{3 +}^{\dagger}a_{4 
+}^{\dagger} \, |0 \rangle \, .
\end{equation}
Using Eqs.~(\ref{output2}) and (\ref{matrix4}), we then find that the collection of one photon per output port prepares the system in the postselected state 
\begin{equation}
|\phi_{\rm pro} \rangle = \sum_{\sigma} \Bigg[ \prod_{i=1}^4 U_{\sigma (i)i} \, b_{i +}^{\dagger} \Bigg] \, 
|0 \rangle = 0 \, .
\end{equation} 
This means, that it is impossible to pass four photons in the same state through the considered setup with each of them leaving the multiport in a different output port. More generally speaking, the state with four photons in the same state does not contribute to the event of collecting one photon per output port. It is therefore impossible to prepare any superposition containing the states $b_{1 +}^{\dagger}b_{2 +}^{\dagger}b_{3 
+}^{\dagger}b_{4 +}^{\dagger} \, |0 \rangle$ or $b_{1 -}^{\dagger}b_{2 -}^{\dagger}b_{3 -}^{\dagger}b_{4 -}^{\dagger} \, |0 \rangle$, respectively. The reason is destructive interference of certain photon states within the linear optics setup, which plays a crucial role for the generation of multiphoton entanglement via postselection. 

\subsection{The 4-photon {\em W}-state} \label{W4}

We now focus our attention on the case, where input port 1 is entered by a photon prepared in $|+ \rangle$ while all other input ports are entered by a photon in $|- \rangle$, i.e.
\begin{equation} \label{Win}
|\phi^W_{\rm in} \rangle = a_{1 +}^{\dagger}a_{2 -}^{\dagger}a_{3 -}^{\dagger}a_{4 -}^{\dagger} \, |0 \rangle \, .
\end{equation}
Using again Eqs.~(\ref{output2}) and (\ref{matrix4}), we find that the collection of one photon per output port
corresponds to the postselected 4-photon state 
\begin{eqnarray} \label{proni}
|\phi_{\rm pro}^W \rangle &=& \sum_{j=1}^4 U_{j1} \, b_{j +}^{\dagger} \, 
\sum_{\sigma_j}  \Bigg[  
\prod_{i=2}^4 U_{\sigma_j(i) i} \, b_{\sigma_j (i) -}^{\dagger} \Bigg] \, 
|0 \rangle \, , \nonumber \\&&
\end{eqnarray} 
where the $\sigma_j$ are the $3!$ permutations that map the list $\{2, \, 3, \,  4\}$ onto the list $\{1, \, ..., \, (j-1), \, (j+1), \, ..., \, 4 \}$. If $|j_{\rm out} \rangle$ denotes the state with one photon in $|+ \rangle$ in output port $j$ and one photon in $|-\rangle$ everywhere else, 
\begin{equation} \label{z}
|j_{\rm out} \rangle \equiv b_{N -}^{\dagger} \, . \, . \, . \, b_{(j+1) -}^{\dagger} b_{j +}^{\dagger} b_{(j-1) -}^{\dagger} \, . \, . \, . \, b_{1 -}^{\dagger} \, |0\rangle \, ,
\end{equation}
and $\beta_j$ is a complex probability amplitude, then the output state (\ref{proni}) can be written as
\begin{eqnarray} \label{pro3}
|\phi_{\rm pro}^W \rangle &=& \sum_{j=1}^4 \beta_j  \, |j_{\rm out} \rangle \, .
\end{eqnarray}
Furthermore, we introduce the reduced transition matrices $U_{\rm red}^{(j)}$, which are obtained by deleting the first column and the $j$-th row of the transition matrix $U$. Then one can express each $\beta_j$ as the permanent \cite{Horn,perms} of a matrix,
\begin{equation} \label{c}
\beta_j = U_{j 1} \sum_{\sigma_j} \prod_{i=2}^4 U_{\sigma_j (i)i} 
= U_{j 1} \, {\rm perm} \, \left( U_{\rm red}^{(j) {\rm T}} \right) \, .
\end{equation}
The output state (\ref{pro3}) equals a {\em W}-state, if the coefficients $\beta_j$ are all of the same size and differ from each other at most by a phase factor. 

To show that this is indeed the case, we calculate the reduced matrices $U_{\rm red}^{(j)}$ explicitly \cite{note2} and obtain 
\begin{eqnarray} \label{sing}
&& \hspace*{-0.5cm} U_{\rm red}^{(1)} = {\textstyle {1 \over 2}} 
\left( \begin{array}{rrrr} \omega_4 & \omega_4^2 & \omega_4^3 \\ \omega_4^2 & \omega_4^4 & 
\omega_4^6 \\ \omega_4^3 & \omega_4^6 & \omega_4^9 \end{array} \right) \, , ~
U_{\rm red}^{(2)} =  {\textstyle {1 \over 2}} 
\left( \begin{array}{rrrr} 1 & 1 & 1 \\ \omega_4^2 & \omega_4^4 & \omega_4^6 \\ \omega_4^3  
& \omega_4^6 & \omega_4^9 \end{array} \right) \, ,  \nonumber \\
&& \hspace*{-0.5cm} U_{\rm red}^{(3)} = {\textstyle {1 \over 2}} 
\left( \begin{array}{rrrr} 1 & 1 & 1 \\ \omega_4 & \omega_4^2 & \omega_4^3 \\ \omega_4^3 & 
\omega_4^6 & \omega_4^9 \end{array} \right) \, , ~
U_{\rm red}^{(4)} =  {\textstyle {1 \over 2}} 
\left( \begin{array}{rrrr} 1 & 1 & 1 \\ \omega_4 & \omega_4^2 & \omega_4^3 \\ \omega_4^2 & 
\omega_4^4 & \omega_4^6 \end{array} \right) \, . \nonumber \\
\end{eqnarray}
The coefficients $\beta_j$ differ at most by a phase factor, if the norm of the permanents of the transpose of these reduced matrices is for all $j$ the same. To show that this is the case we now define the vector 
\begin{equation}
{\bf v} = (\omega_4, \omega_4^2, \omega_4^3) \, ,
\end{equation}
multiply each row of the matrix $U_{\rm red}^{(1)}$ exactly $(j-1)$ times with ${\bf v}$ and obtain the new matrices   
\begin{eqnarray}
&&  \hspace*{-0.5cm} \tilde{U}_{\rm red}^{(1)} = U_{\rm red}^{(1)} \, , ~
\tilde{U}_{\rm red}^{(2)} =  {\textstyle {1 \over 2}}
\left( \begin{array}{rrrr} \omega_4^2 & \omega_4^4 & \omega_4^6 \\ \omega_4^3 & \omega_4^6 & \omega_4^9 \\ 1 & 1 & 1 \end{array} \right) \, , ~ \nonumber \\
&& \hspace*{-0.5cm} \tilde{U}_{\rm red}^{(3)} =  {\textstyle {1 \over 2}}
\left( \begin{array}{rrrr} \omega_4^3 & \omega_4^6 & \omega_4^9 \\ 1 & 1 & 1 \\ \omega_4 &
\omega_4^2 & \omega_4^3 \end{array} \right) \, , ~ 
\tilde{U}_{\rm red}^{(4)} = U_{\rm red}^{(4)}  \, . 
\end{eqnarray}
The above described multiplication amounts physically to the multiplication of the photon input state with an overall phase factor and 
\begin{equation}
\left| \, {\rm perm} \left( U_{\rm red}^{(1) {\rm T}} \right) \, \right| = \left| \, {\rm perm} \left( \tilde{U}_{\rm red}^{(j) {\rm T}} \right) \, \right| \, . 
\end{equation}
Moreover, using the cyclic symmetry of permanents \cite{Horn}, we see that 
\begin{equation}
{\rm perm} \left( U_{\rm red}^{(j) {\rm T}} \right) = {\rm perm} \left( \tilde{U}_{\rm red}^{(j) {\rm T}} \right) \, .
\end{equation}
This implies together with Eq.~(\ref{c}) that the norm of the coefficients $\beta_j$ is indeed the same for all $j$. Furthermore, using the above argument based on the multiplication of phase factors to the photon input state, one can show that
\begin{equation} \label{phase}
\beta_j= \beta_1 \left( \prod_{k=0}^3 \omega_4^k \right)^{j-1} \, .
\end{equation}
Inserting this into Eq.~(\ref{proni}), we find that the postselected state with one photon per output port equals, after normalisation \cite{note}, the {\em W}-state 
\begin{eqnarray} \label{beta}
|\hat{\phi}_{\rm pro}^W \rangle 
&=& {\textstyle {1 \over 2}} \, \big[ \, b_{1 +}^{\dagger}b_{2 -}^{\dagger}b_{3 
-}^{\dagger}b_{4 -}^{\dagger}-b_{1 -}^{\dagger}b_{2 
+}^{\dagger}b_{3 -}^{\dagger}b_{4 -}^{\dagger} \nonumber \\
&& + b_{1 -}^{\dagger} b_{2 -}^{\dagger} b_{3 +}^{\dagger} b_{4 -}^{\dagger} 
-b_{1 -}^{\dagger} b_{2 -}^{\dagger} b_{3 -}^{\dagger} b_{4  
+}^{\dagger} \, \big] \, |0 \rangle \, .~~ \nonumber \\&&
\end{eqnarray} 
In analogy, we conclude that an input state with one photon in $|-\rangle$ in input port 1 and a photon in $|+ \rangle$ in each of the other input ports, results in the preparation of the {\em W}-state 
\begin{eqnarray} \label{beta2}
|\hat{\phi}_{\rm pro}^{W'} \rangle 
&=&  {\textstyle {1 \over 2}} \, \big[ \, b_{1 -}^{\dagger}b_{2 +}^{\dagger}b_{3 
+}^{\dagger}b_{4 +}^{\dagger}-b_{1 +}^{\dagger}b_{2 -}^{\dagger}b_{3 
+}^{\dagger}b_{4 +}^{\dagger} \nonumber \\
&& +b_{1 +}^{\dagger}b_{2 +}^{\dagger}b_{3 -}^{\dagger}b_{4 +}^{\dagger}-b_{1 
+}^{\dagger}b_{2 +}^{\dagger}b_{3 +}^{\dagger}b_{4 -}^{\dagger} \, \big] \, 
|0 \rangle  \nonumber \\&&
\end{eqnarray}
under the condition of the collection of one photon per output port. Both states, (\ref{beta}) and (\ref{beta2}), can be generated with probability 
\begin{equation} \label{s}
P_{\rm suc}={\textstyle {1 \over 16}} \, .
\end{equation}
Transforming them into the usual form of a $W$-state with equal coefficients of all amplitudes \cite{duer} only requires further implementation of a Pauli $\sigma_z$ operation (i.e.~a state dependent sign flip) on either the first and the third or the second and the fourth output photon, respectively.

\subsection{The 4-photon GHZ-state} \label{fun}

Besides generating {\em W}-states, the proposed setup can also be used to prepare 4-photon GHZ-states. This requires, feeding each of the input ports 1 and 3 with one photon in $|+ \rangle$ while the input ports 2 and 4 should each be entered by a photon in $|- \rangle$ such that
\begin{equation}
|\phi_{\rm in}^{\rm GHZ} \rangle = a_{1 +}^{\dagger}a_{2 -}^{\dagger}a_{3 +}^{\dagger}a_{4 -}^{\dagger} \, |0 \rangle \, .
\end{equation}
Calculating again the output state under the condition of collecting one photon per output port, we obtain 
\begin{eqnarray} 
|\phi_{\rm pro}^{\rm GHZ} \rangle 
&=& \sum_{\sigma}U_{\sigma (1)1}U_{\sigma (2)2}U_{\sigma (3)3}U_{\sigma (4)4} 
\nonumber \\
&& b_{\sigma (1)+}^{\dagger}b_{\sigma (2) -}^{\dagger}b_{\sigma (3) +}^{\dagger}b_{\sigma (4) -}^{\dagger} \, |0 \rangle \, , ~~
\end{eqnarray} 
where the $\sigma$ are the $4!$ permutations that map the list $\{1, \, 2, \, 3, \, 4 \}$ onto itself. On simplification, one finds that there are only two constituent states with non-zero coefficients and $|\phi_{\rm pro}^{\rm GHZ} \rangle$ becomes after normalisation
\begin{eqnarray} \label{outGHZ}
|\hat{\phi}_{\rm pro}^{\rm GHZ} \rangle = {\textstyle {1 \over \sqrt{2}}} \, \big[ \, b_{1 +}^{\dagger}b_{2 
-}^{\dagger}b_{3 +}^{\dagger}b_{4 -}^{\dagger}-b_{1 -}^{\dagger}b_{2+}^{\dagger}b_{3 -}^{\dagger}b_{4 +}^{\dagger} \, \big] \, |0 \rangle \, , \nonumber \\&&
\end{eqnarray} 
which equals the GHZ-state up to local operations. Transforming (\ref{outGHZ}) into the usual form of the GHZ-state requires changing the state of two of the photons, for example, from $|+ \rangle$ into $|- \rangle$. This can be realised by applying a Pauli $\sigma_x$ operation to the first output port as well as a $\sigma_y$ operation to the third output. 

Finally, we remark that the probability for the creation of the GHZ-state (\ref{ss}) is twice as high as the probability for the generation of a {\em W}-state (\ref{s}), 
\begin{equation} \label{ss}
P_{\rm suc} = {\textstyle {1 \over 8}} \, .
\end{equation} 
Unfortunately, the experimental setup shown in Fig.~\ref{scheme} does not allow for the preparation of GHZ-states for arbitrary photon numbers $N$. For a detailed description of polarisation entangled GHZ states using a different network of $50:50$ and polarising beam splitters see Ref.~\cite{Sagi}.

\subsection{The 4-photon double singlet state}

For completeness, we now ask for the output of the proposed state preparation scheme, given that the input state equals
\begin{equation}
|\phi_{\rm in} \rangle = a_{1 +}^{\dagger} a_{2 +}^{\dagger} a_{3 -}^{\dagger} a_{4 -}^{\dagger} |0 \rangle \, .
\end{equation}
Proceeding as above, we find that this results in the preparation of the state
\begin{eqnarray}
|\phi_{\rm pro}^{\rm DS} \rangle
&=& \sum_{\sigma}U_{\sigma (1)1}U_{\sigma (2)2}U_{\sigma (3)3}U_{\sigma (4)4} 
\nonumber \\
&& b_{\sigma (1)+}^{\dagger}b_{\sigma (2) +}^{\dagger}b_{\sigma (3)-}^{\dagger}b_{\sigma (4)-}^{\dagger}\, |0 \rangle 
\end{eqnarray} 
under the condition of the collection of one photon per output port. Here the permutation operators $\sigma$ are defined as in Section \ref{fun}, which yields
\begin{eqnarray} \label{DS}
|\hat{\phi}_{\rm pro}^{\rm DS}  \rangle 
&=& {\textstyle {1 \over 2}} \, \big[ \, b_{1 +}^{\dagger}b_{2 +}^{\dagger}b_{3 
-}^{\dagger}b_{4 -}^{\dagger}+b_{1 -}^{\dagger}b_{2 
-}^{\dagger}b_{3 +}^{\dagger}b_{4 +}^{\dagger} \nonumber \\
&& -b_{1 +}^{\dagger}b_{2 -}^{\dagger}b_{3 -}^{\dagger}b_{4 +}^{\dagger}-b_{1 
-}^{\dagger}b_{2 +}^{\dagger}b_{3 +}^{\dagger}b_{4 -}^{\dagger} \, ] \, |0 \rangle \, . ~~~~
\end{eqnarray} 
This state can be prepared with probability
\begin{equation}
P_{\rm suc} = {\textstyle {1 \over 16}} \, . 
\end{equation}
The state (\ref{DS}) is a double singlet state, i.e.~a tensor product of two 2-photon singlet states, with a high robustness against decoherence \cite{Weinfurter}.

\subsection{The general 4-photon case}

Finally, we consider the situation where the input state is of the general form (\ref{in}). Calculating Eq.~(\ref{output2}), we find that the unnormalised output state under the condition of one photon per output port equals in this case
\begin{eqnarray} \label{final}
|\phi_{\rm pro} \rangle &=&
{\textstyle {{\rm i} \over 4}} \big( \, \gamma_1+\gamma_2-\gamma_3-\gamma_4 \, 
\big) \, |\hat{\phi}_{\rm pro}^{\rm DS} \rangle \nonumber \\
&& + {\textstyle {1 \over 2 \sqrt{2}}} \big( \, \gamma_5-\gamma_6\, \big) \, 
|\hat{\phi}_{\rm pro}^{\rm GHZ} \rangle \nonumber \\
&& + {\textstyle {1 \over 4}} \big( \, \gamma_8 +\gamma_{10}-\gamma_7-\gamma_9 
\, \big) \, |\hat{\phi}_{\rm pro}^{W} \rangle \nonumber \\
&& + {\textstyle{1 \over 4}} \big( \, 
\gamma_{12}+\gamma_{14}-\gamma_{11}-\gamma_{13} \, \big) \, |\hat{\phi}_{\rm 
pro}^{W'} \rangle ~~~
\end{eqnarray} 
with the coefficients
\begin{eqnarray} \label{coefficients}
&\gamma_{1}=\alpha_{1+}\alpha_{2+}\alpha_{3-}\alpha_{4-} \, , ~
& \gamma_{2}=\alpha_{1-}\alpha_{2-}\alpha_{3+}\alpha_{4+}  \, ,\nonumber \\
& \gamma_{3}=\alpha_{1-}\alpha_{2+}\alpha_{3+}\alpha_{4-} \, , ~
& \gamma_{4}=\alpha_{1+}\alpha_{2-}\alpha_{3-}\alpha_{4+}  \, ,\nonumber \\
& \gamma_{5}=\alpha_{1+}\alpha_{2-}\alpha_{3+}\alpha_{4-} \, , ~
& \gamma_{6}=\alpha_{1-}\alpha_{2+}\alpha_{3-}\alpha_{4+}  \, , \nonumber \\
& \gamma_{7}=\alpha_{1+}\alpha_{2-}\alpha_{3-}\alpha_{4-} \, , ~
& \gamma_{8}=\alpha_{1-}\alpha_{2+}\alpha_{3-}\alpha_{4-}  \, , \nonumber \\
& \gamma_{9}=\alpha_{1-}\alpha_{2-}\alpha_{3+}\alpha_{4-} \, , ~
& \gamma_{10}=\alpha_{1-}\alpha_{2-}\alpha_{3-}\alpha_{4+}  \, , \nonumber \\
& \gamma_{11}=\alpha_{1-}\alpha_{2+}\alpha_{3+}\alpha_{4+} \, , ~
& \gamma_{12}=\alpha_{1+}\alpha_{2-}\alpha_{3+}\alpha_{4+}  \, , \nonumber \\
& \gamma_{13}=\alpha_{1+}\alpha_{2+}\alpha_{3-}\alpha_{4+} \, , ~
& \gamma_{14}=\alpha_{1+}\alpha_{2+}\alpha_{3+}\alpha_{4-} \, .~~~~~~
\end{eqnarray} 
The form of the coefficients (\ref{coefficients}) reflects the full symmetry of the transformation of the input state. Each of the entangled states $|\hat{\phi}_{\rm pro}^{\rm DS} \rangle$, $|\hat{\phi}_{\rm pro}^{\rm GHZ} \rangle$, $|\hat{\phi}_{\rm pro}^{W} \rangle$ and $|\hat{\phi}_{\rm pro}^{W'} \rangle$ are generated independently from the different constituent parts of the input (\ref{in}). Besides, Eq.~(\ref{final}) shows that the output state is constrained to be of a certain symmetry, namely the symmetry introduced by the symmetry of the $N \times N$ Bell multiport and the postselection criteria of finding one photon per output port.

\section{The generation of $N$-photon {\em W}-states} \label{doubleW} 

Using the same arguments as in Section \ref{W4}, we now show that the $N \times N$ Bell multiport beam splitter can be used for the generation of {\em W}-states for arbitrary photon numbers $N$. Like Bell states, {\em W}-states are highly entangled but their entanglement is more robust \cite{duer}. Moreover, as $N$ increases, {\em W}-states perform better than the corresponding GHZ states against noise admixture in experiments to violate local realism \cite{Sen(De)} and are important for optimal cloning protocols \cite{buzek}. 

In analogy to Eq.~(\ref{Win}), we assume that the initial state contains one photon in $|+ \rangle$ in the first input port, while every other input port is entered by a photon prepared in $|- \rangle$ so that
\begin{eqnarray} 
|\phi_{\rm in} \rangle &=& a_{1 +}^{\dagger} \prod_{i=2}^N a_{i -}^{\dagger} \, 
|0 \rangle \, . 
\end{eqnarray} 
Using Eq.~(\ref{output2}), we find that the state of the system under the condition of the collection of one photon per output port equals
\begin{eqnarray} \label{output3} 
|\phi_{\rm pro} \rangle &=& \sum_{j=1}^N U_{j1} \, b_{j +}^{\dagger} \, 
\sum_{\sigma_j}  \Bigg[  
\prod_{i=2}^N U_{\sigma_j(i) i} \, b_{\sigma_j (i) -}^{\dagger} \Bigg] \, 
|0 \rangle \, , \nonumber \\&& 
\end{eqnarray} 
where the $\sigma_j$ are the $(N-1)!$ permutations that map the list $\{2, \, 3, \, ..., \, N\}$ onto the list $\{1,\, 2, \, ..., \, (j-1), \, (j+1), \, ..., \, N \}$. As expected, the output is a superposition of all states with one photon in $|+ \rangle$ and all other photons prepared in $|- \rangle$. 

To prove that Eq.~(\ref{output3}) describes indeed a {\em W}-state, we use again the notation introduced in Eqs.~(\ref{z}) and (\ref{pro3}) and write 
\begin{eqnarray} \label{pro4} 
|\phi_{\rm pro} \rangle & \equiv& \sum_j \beta_j  \, |j_{\rm out} \rangle \, .
\end{eqnarray} 
To show that the coefficients $\beta_j$ differ from $\beta_1$ at most by a phase factor, we express the amplitudes $\beta_j$ as in Eq.~(\ref{c}) using the permanents of the reduced transition matrices and find
\begin{equation} \label{ccc} 
\beta_j = U_{j 1} \sum_{\sigma_j} \prod_{i=2}^N U_{\sigma_j (i)i} = U_{j 1} \, {\rm perm} \, \left( U_{\rm red}^{\rm T}(j) \right) \, . 
\end{equation} 
Inserting the concrete form of the transition matrix $U$, this yields 
\begin{equation} 
\beta_j = \frac{1}{\sqrt{N^N}}\sum_{\sigma_j} \prod_{i=2}^N \omega_N^{(\sigma_j 
(i)-1)(i-1)} \, .~~ 
\end{equation} 
Proceeding as in Section \ref{W4}, we now multiply $\beta_j$ with the phase factor
\begin{equation} 
v_j \equiv \left( \prod_{k=0}^{N-1}  \omega_N^k \right)^{-(j-1)} 
\end{equation} 
and obtain
\begin{eqnarray} \label{generalproof} 
v_j \, \beta_j &=& {\textstyle {1 \over \sqrt{N^N}}} \sum_{\sigma_j} 
\prod_{i=2}^N \omega_N^{(\sigma_j (i)-j)(i-1)} \nonumber \\
&=&  {\textstyle {1 \over \sqrt{N^N}}} \sum_{\sigma_j} \prod_{i=2}^N 
\omega_N^{\big({\rm mod}_N (\sigma_j (i)-j) \big)(i-1)} \, .~~
\end{eqnarray} 
The expression ${\rm mod}_N(\sigma_j (i)-j)+1$ represents a set of $(N-1)!$ permutations that map $\{2, \, 3, \, ..., \, N\}$ onto the list $\{2, \, 3, \, ..., \, N\}$. It is therefore equivalent to the permutations $\sigma_1(i)$, which allows us to simplify Eq.~(\ref{generalproof}) even further and to show that
\begin{equation} \label{close} 
v_j \, \beta_j = {\textstyle {1 \over \sqrt{N^N}}} 
\sum_{\sigma_1} \prod_{i=2}^N \omega_N^{(\sigma_1 (i)-1)(i-1)} = \beta_1 \, .~~ 
\end{equation} 
From this and the fact that $1+2+...+(N-1)={1 \over 2} N(N-1)$ we finally arrive at the relation 
\begin{eqnarray} \label{relative} 
\beta_j &=& \left( \prod_{k=0}^{N-1} \omega_N^k \right)^{j-1} \beta_1 \nonumber \\ 
&=& \left\{ \begin{array}{cl} \beta_1 \, , & {\rm if}~N~{\rm is~odd} \, , \\  
(-1)^{j-1} \, \beta_1 \, , & {\rm if}~N~{\rm is~even} \, . \end{array} \right. 
\end{eqnarray} 
This shows that the amplitudes $\beta_j$ are all of the same size and the Bell multiport can indeed generate $N$-photon {\em W}-states. If one wants the coefficients $\beta_j$ to be exactly the same, one can remove unwanted minus signs in case of even photon numbers by applying a $\sigma_z$ operation in each output port with an even number $j$. 

In the case $N=2$, the above described state preparation scheme reduces to the familiar example, where two photons prepared in the two orthogonal states $|+ \rangle$ and $|- \rangle$ pass through a 50:50 beam splitter. The collection of one photon in the each output port prepares the system in this case in the state ${1 \over \sqrt{2}} \, [ \, b_{1+}^\dagger b_{2-}^\dagger - b_{1-}^\dagger b_{2+}^\dagger \, ] \, |0 \rangle$, which can be transformed into ${1 \over \sqrt{2}} \, [ \, b_{1+}^\dagger b_{2-}^\dagger + b_{1-}^\dagger b_{2+}^\dagger \, ] \, |0 \rangle$ by performing a conditional sign flip, i.e.~depending on whether the photon is in $|+ \rangle$ or $|-\rangle$, in one of the output ports. 

\subsection{Success probabilities}

\begin{figure}
\begin{minipage}{\columnwidth}
\begin{center}
\resizebox{\columnwidth}{!}{\rotatebox{0}{\includegraphics{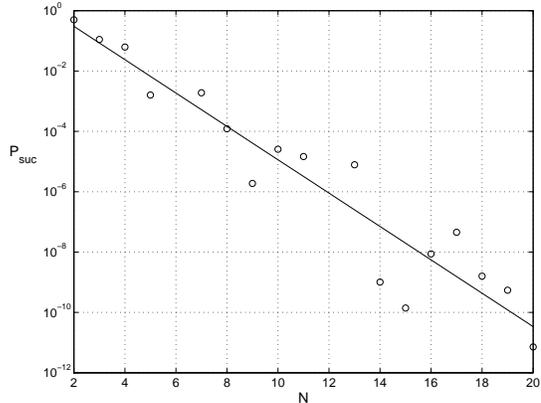}}}  
\end{center}
\caption{The success rate for the generation of $N$-photon {\em W}-states 
$P_{\rm suc}$ as a function of $N$. The solid line approximates the exact 
results via the equation $P_{\rm suc}={\rm e}^{a-bN}$ with $a=1.35 \pm 1.32$ and 
$b=1.27 \pm 0.10$} \label{fireplot}
\end{minipage}
\end{figure}

Let us finally comment on the success rate of the proposed {\em W}-state preparation scheme. Computing the probability (\ref{suc}) can be done by finding the amplitude $\beta_1$ with the help of Eq.~(\ref{ccc}). Although the definition of the permanent of a matrix resembles the definition of the determinant, there exist only few theorems that can be used to simplify their calculation \cite{Horn}. In fact, the computation of the permanent is an NP-complete problem. We~therefore calculated $P_{\rm suc}$ numerically (see Fig.~\ref{fireplot}). 

As it applies to linear optics schemes in general, the success probability decreases unfavourably as the number of qubits involved increases. Here the probability for the collection drops on average exponentially. We observe the interesting effect of a non-monotonic decreasing success probability as $N$ increases. For example, the 
probabilty of success for $N=13$ is higher than for $N=9$. Moreover, for $N=6$ and $N=12$, {\em W}-state generation is not permitted due to destructive interference. This does not apply to $N=18$ which is also a 
multiple of $6$. 

\section{Conclusions} \label{conclusions}

We analysed the generation of multiphoton entanglement with the help of interference and postselection in a linear optics network consisting of an $N \times N$ Bell multiport beam splitter. Each input port should be entered by a single photon prepared in a certain state $|\lambda_i \rangle$. As long as the photons are the same with respect to all other degrees of freedom and it can be guaranteed that photons prepared in the same state overlap within their coherence time inside the linear optics network, the described scheme can be implemented using only a single photon source  \cite{Kuhn2,Mackeever,Lange04,Yamamoto,Weinfurter00,Grangier}. We believe that the described approach allows to entangle much higher photon numbers than what can be achieved in parametric down conversion experiments. 

An, in general, highly entangled output state is obtained under the condition of the collection of one photon per output port. The motivation for this postselection criteria is that distinguishing this state from other output states does not require photon number resolving detectors. Moreover, the photons can easily be processed further and provide a resource for linear optics quantum computing and quantum cryptographic protocols. 

First we analysed the case $N=4$ and showed that the $4 \times 4$ Bell multiport allows for the creation of a variety of highly-symmetric entangled states including the {\em W}-state, the GHZ-state and double singlet states. It was found that some states are easier to prepare than others. A straightforward generalisation of the 4-photon case, yields a scheme for the creation of $N$-photon {\em W}-states. We calculated the rates for successful state preparations and showed that they decrease in a non-monotonic fashion and on average exponentially with $N$.

The motivation for considering a Bell multiport beam splitter was that it only redirects the photons without affecting their inner degrees of freedom. The proposed setup can therefore be used to produce polarisation, time-bin and frequency entanglement, respectively. To generate, for example, polarisation entangled photons, the initial photon states may differ in polarisation but should otherwise be exactly the same. The high symmetry of the Bell multiport beam splitter allows for the generation of a variety of highly entangled symmetric states. To verify the generation of a certain type of entanglement one could, for example, use local measurements as it has recently been proposed by Toth and G\"uhne \cite{Toth}. 

\begin{acknowledgements}
Y. L. L. acknowledges financial support by the DSO National Laboratories in Singapore. A. B. thanks the Royal Society and the GCHQ for funding as a James Ellis University Research Fellow. This work was supported in part by the European Union and the UK Engineering and Physical Sciences Research Council.
\end{acknowledgements}


\begin{references}
\bibitem{Gottesman}
D. Gottesman and I. L. Chuang, Nature {\bf 402}, 390 (1999). 
%
\bibitem{Knill}
E. Knill, R. Laflamme, and G. J. Milburn, Nature {\bf  409}, 46 (2001).
%
\bibitem{Bennett} 
C. H. Bennett and G. Brassard, in {\em Proceedings of the IEEE International 
Conference on Computers, Systems and Signal Processing, Bangalore, India} (IEEE, 
New York, 1984), pp. 175-179.
%
\bibitem{Ekert}
A. K. Ekert, Phys. Rev. Lett. {\bf 67}, 661 (1991).
%
\bibitem{Bell}
J. S. Bell, Physics {\bf 1}, 195 (1965).
%
\bibitem{Aspect}
A. Aspect, P. Grangier, and G. Roger, Phys. Rev. Lett. {\bf 49}, 91 (1982).
%
\bibitem{downconversion}
P. G. Kwiat, K. Mattle, H. Weinfurter, A. Zeilinger, A. V. Sergienko, and Y. Shih, Phys. Rev. Lett. {\bf  75}, 4337 (1995).
%
\bibitem{Gisin} 
J. Brendel, N. Gisin, W. Tittel, and H. Zbinden, Phys. Rev. Lett. {\bf 82}, 2594 (1999);
R. T. Thew, S. Tanzilli, W. Tittel, H. Zbinden, and N. Gisin, Phys. Rev. A {\bf 66}, 062304 (2002); 
H. de Riedmatten, I. Marcikic, W. Tittel, H. Zbinden, D. Collins, and N. Gisin, Phys. Rev. Lett {\bf 92}, 047904 (2004).
%
\bibitem{Gheri}
K. M. Gheri, C. Saavedra, P. T\"{o}rm\"{a}, J. I. Cirac, and P. Zoller, Phys. Rev. A {\bf 58}, R2627 (1998).
%
\bibitem{Lange}
W. Lange and H. J. Kimble, Phys. Rev. A {\bf 61}, 063817 (2000).
%
\bibitem{schoen}
C. Sch\"on, E. Solano, F. Verstraete, J. I. Cirac, and M. M. Wolf, {\em sequential generation of entangled multi-qubit states}, quant-ph/0501096.
%
\bibitem{jpa}
Y. L. Lim and A. Beige, J. Phys. A {\bf 38}, L7 (2005).
%
\bibitem{orlando}
Y. L. Lim and A. Beige, Proc. SPIE {\bf 5436}, 118 (2004).
%
\bibitem{moonlight}
Y. L. Lim, A. Beige, and L. C. Kwek, {\em Quantum computing with distant single photon sources with insurance}, quant-ph/0408043.
%
\bibitem{Sipe}
G. G. Lapaire, P. Kok, J. P. Dowling, and J. E. Sipe, Phys. Rev. A {\bf 68}, 042314 (2003). 
%
\bibitem{Shih}
Y. H. Shih and C. O. Alley, Phys. Rev. Lett. {\bf 61}, 2921 (1988).
%
\bibitem{recent}
D. Fattal, K. Inoue, J. Vuckovic, C. Santori, G. S. Solomon, and Y. Yamamoto, Phys. Rev. Lett. {\bf 92}, 037903 (2004).
%
\bibitem{Weinfurter}
M. Eibl, S. Gaertner, M. Bourennane, C. Kurtsiefer, M. Zukowski, and H. Weinfurter, Phys. Rev. Lett. {\bf 90}, 200403 (2003);
M. Bourennane, M. Eibl, S. Gaertner, C. Kurtsiefer, A. Cabello, and H. Weinfurter, Phys. Rev. Lett. {\bf 92}, 107901 (2004). ÊÊ
%
\bibitem{pan}
Z. Zhao, T. Yang, Y.-A. Chen, A.-N. Zhang, M. Zukowski, and J.-W. Pan, Phys. Rev. Lett. {\bf 91}, 180401 (2003);
Z. Zhao, A.-N. Zhang, Y.-A. Chen, H. Zhang, J.-F. Du, T. Yang, and J.-W. Pan, Phys. Rev. Lett. {\bf 94}, 030501 (2005).
%
\bibitem{Law} 
C. K. Law and H. J. Kimble, J. Mod. Optics {\bf 44}, 2067 (1997). 
%
\bibitem{Kuhn1} 
A. Kuhn, M. Hennrich, T. Bondo, and G. Rempe, Appl. Phys. B {\bf 69}, 373 (1999). 
%
\bibitem{Duan} 
L. -M. Duan, A. Kuzmich, and H. J. Kimble, Phys. Rev. A {\bf 67}, 032305 (2003). 
%
\bibitem{Kwiat}
E. Jeffrey, N. A. Peters, P. G. Kwiat, and G. Kwiat, New J. Phys.  {\bf 6}, 100 (2004).
%
\bibitem{Kuhn2} 
M. Hennrich, T. Legero, A. Kuhn, and G. Rempe, Phys. Rev. Lett. {\bf 85}, 4872 (2000); 
A. Kuhn, M. Hennrich and G. Rempe, Phys. Rev. Lett. {\bf 89}, 067901 (2002). 
%
\bibitem{Mackeever} 
J. McKeever, A. Boca, A. D. Boozer, R. Miller, J. R. Buck, A. Kuzmich and H. J. Kimble, Science {\bf 303}, 1992 (2004). 
%
\bibitem{Lange04} 
M. Keller, B. Lange, K. Hayasaka, W. Lange, and H. Walther, Nature {\bf 431}, 1075 (2004). 
%
\bibitem{Yamamoto} 
O. Benson, C. Santori, M. Pelton, and Y. Yamamoto, Phys. Rev. Lett. {\bf 84}, 2513 (2000);
M. Pelton, C. Santori, J. Vuckovic, B. Zhang, G. S. Solomon, J. Plant, and Y. Yamamoto, Phys. Rev. Lett {\bf 89}, 233602 (2002).
%
\bibitem{Weinfurter00} 
C. Kurtsiefer, S. Mayer, P. Zarda, and H. Weinfurter, Phys. Rev. Lett. {\bf 85}, 290 (2000). 
%
\bibitem{Grangier} 
A. Beveratos, R. Brouri, T. Gacoin, A. Villing, J-P. Poizat, and P. Grangier, Phys. Rev. Lett. {\bf 89}, 187901 (2002). 
%
\bibitem{Zukowski}
M. Zukowski, A. Zeilinger, and M. A. Horne, Phys. Rev. A {\bf 55}, 2564 (1997).
%
\bibitem{Kok}
H. Lee, P. Kok, N. J. Cerf, and J. P. Dowling, Phys. Rev. A {\bf 65}, 030101(R) (2002).
%
\bibitem{Fiurasek}
J. Fiurasek, Phys. Rev. A {\bf 65}, 053818 (2002).
%
\bibitem{Xubo}
X. B. Zou, K. Pahlke, and W. Mathis, Phys. Rev. A {\bf 66}, 014102 (2002). 
%
\bibitem{Pryde}
G. J. Pryde and A. G. White, Phys. Rev. A {\bf 68}, 052315 (2003).
%
\bibitem{Wang}
X.-B. Wang, Phys. Rev. A {\bf 68}, 042304 (2003).
%
\bibitem{Sagi}
Y. Sagi, Phys. Rev. A {\bf 68}, 042320 (2003).
%
\bibitem{Cerf}
J. Fiurasek, S. Massar, and N. J. Cerf, Phys. Rev. A {\bf 68}, 042325 (2003). 
%
\bibitem{Franson}
T. B. Pittman, B. C. Jacobs, and J. D. Franson, Phys. Rev. Lett. {\bf 88}, 257902 (2002).  
%
\bibitem{Paivi1}
P. T\"{o}rm\"{a}, S. Stenholm, and I. Jex, Phys. Rev. A {\bf 52}, 4853 (1995).
%
\bibitem{tb}
N. Gisin, G. Ribordy, W. Tittel, and H. Zbinden, Rev. Mod. Phys. {\bf 74}, 145 (2002).
%
\bibitem{Paivi2}
P. T\"{o}rm\"{a}, Phys. Rev. Lett {\bf 81}, 2185 (1998).
%
\bibitem{Horn}
R. A. Horn and C. R. Johnson, {\em Matrix Analysis}, Cambridge University Press (1985).
%
\bibitem{perms}
S. Scheel, {\em Permanents in linear optics networks}, quant-ph/0406127.
%
\bibitem{note2}
The reason for not simplifying these matrices is that the following equations provide the motivation for the proof of the general case in Section \ref{doubleW}.
%
\bibitem{note}
In the following we denote normalised states by marking them with the $\hat{~}$ symbol.
%
\bibitem{duer}
W. D\"ur, G. Vidal, and J. I. Cirac, Phys. Rev. A {\bf 62}, 062314 (2000). 
%
\bibitem{Sen(De)}
A. Sen(De), U. Sen, M. Wie{\'s}niak, D. Kaszlikowski, and M. Zukowski, Phys. Rev. A {\bf 68}, 062306 (2003).
%
\bibitem{buzek}
V. Bu$\check{\rm z}$ek and M. Hillery, Phys. Rev. A {\bf 54}, 1844 (1996).
%
\bibitem{Toth}
G. Toth and O. G\"uhne, {\em Detecting genuine multipartite entanglement with local measurements}, quant-ph/0405165.
\end{references}
\end{document}